\documentclass[reprint,3p]{elsarticle}

\journal{Physica B: Condensed Matter}

\usepackage[english]{babel}
\usepackage{latexsym}
\usepackage{psfrag}
\usepackage{amsmath}
\usepackage{amssymb}
 
\renewcommand{\v}[1]{\ensuremath{\mathbf{#1}}} 
\newcommand{\gv}[1]{\ensuremath{\mbox{\boldmath$ #1 $}}} 
\renewcommand{\d}[2]{\frac{d #1}{d #2}} 
\newcommand{\dd}[2]{\frac{d^2 #1}{d #2^2}} 
\let\baraccent=\= 
\renewcommand{\=}[1]{\stackrel{#1}{=}} 

\newcommand{\vnabla}{\gv{\nabla}}

\providecommand{\vr}{\v{r}}
\renewcommand{\vr}{\v{r}}

\newcommand{\vk}{\v{k}}
\newcommand{\vG}{\v{G}}

\newcommand{\vp}{\v{p}}

\newcommand{\rd}{{\rm d}}
\newcommand{\im}{{\rm i}}

\usepackage{graphicx}
\graphicspath{./}

\begin{document}

\renewcommand*{\today}{April 16, 2015}

\DeclareGraphicsExtensions{.eps}

\begin{frontmatter}
    
\title{An envelope function formalism for lattice-matched heterostructures}

\author[ua,imec]{Maarten~Van~de~Put}
\ead{maarten.vandeput@imec.be}
\author[utd]{William~Vandenberghe}
\author[ua,imec]{Wim~Magnus}
\author[ua,imec]{Bart~Sor\'ee}

\address[ua]{Department~of~Physics, Universiteit~Antwerpen, Antwerpen, Belgium}
\address[imec]{imec, Kapeldreef~75, 3001~Heverlee, Belgium}
\address[utd]{Department~of~Materials~Science~and~Engineering, The~University~of~Texas~at~Dallas, Dallas, USA}

\begin{abstract}
The envelope function method traditionally employs a single basis set which, in practice, relates to a single material because the $\vk\cdot\vp$ matrix elements are generally only known in a particular basis. 
In this work, we defined a basis function transformation to alleviate this restriction. 
The transformation is completely described by the known inter-band momentum matrix elements. 
The resulting envelope function equation can solve the electronic structure in lattice matched heterostructures without resorting to boundary conditions at the interface between materials, while all unit-cell averaged observables can be calculated as with the standard envelope function formalism. 
In the case of two coupled bands, this heterostructure formalism is equivalent to the standard formalism while taking position dependent matrix elements.
\end{abstract}


\end{frontmatter}

\section{Introduction}

To calculate the band structure of semiconductors, a number of methods
are available: the ${\bf k}\cdot{\bf p}$ method\cite{Cardona:1966ki},
the tight-binding method\cite{boykin2004valence} and the pseudopotential
method\cite{chelikowsky1976nonlocal}. The latter two methods are
atomistic while the ${\bf k}\cdot{\bf p}$ method is not and will
be computationally more efficient when atomic resolution is not required.
The ${\bf k\cdot{\bf p}}$ method has been successfully used to explain
many physical phenomena such as band-to-band tunneling (BTBT)\cite{kane1960zener,vandenberghe2010zener}
or topological insulators\cite{bernevig2006quantum}. The ${\bf k}\cdot{\bf p}$
method has also been used successfully to study tunnel field-effect
transistors (TFETs)\cite{conzatti2012strain}. But with a recent interest
in TFETs composed of different materials\cite{vandenberghe2012model},
the need for a ${\bf k}\cdot{\bf p}$-based framework that can deal
with heterostructures emerges.

Historically, the $\vk\cdot\vp$ formalism for bulk has been extended in various ways to an envelope function formalism which can provide the electronic structure for arbitrary potentials.
The rigorous derivation of an exact envelope function formalism by Burt\cite{Burt:1988fa} provides a solid physical and mathematical basis for most extensions of the $\vk\cdot\vp$ formalism.
The exact envelope function equations described in Burt's work involve a non-local interaction of the external potential, but under the right approximations (a slowly varying potential) they reproduce local equations. 
In literature the local equations are used almost exclusively because they are numerically manageable due to known, empirically determined, matrix elements while Burt's exact equations require knowledge of the complete basis set.

Several authors have proposed strategies to adopt the local envelope function formalism to heterostructures, we recognize three common approaches.
The first approach correctly accounts for the different material basis sets by relying on a complete knowledge of the basis functions\cite{Burt:1992,Burt:1999cw}. 
This method is a nice theoretical exercise, but highly impractical, as most benefits of the envelope function are lost when full knowledge of the basis functions is required.
A second strategy (implicitly) assumes the basis functions for different materials used for the envelope function expansion to be identical for different materials, which enables bulk-like envelope function expansion. However, in practice the matrix elements are often taken to be the local material specific bulk matrix elements in their natural basis\cite{BoyerRichard:2011hg,ElKurdi:2006ul}, which is inconsistent with them being identical.
Foreman details a third strategy, he accounts for the difference in basis functions by introducing a interface term to describe additional couplings between the two materials, while using material-dependent basis functions throughout.\cite{Foreman:1996gt} This last strategy has recently been employed in the context of studying the homogenization limit for heterostructure $\vk\cdot\vp$ multi-band models and the derivation of an optimal effective mass model for heterostructures\cite{morandi:2014}.

In this paper we introduce a new set of envelope function equations which is capable of describing the electronic structure in lattice-matched heterostructures, taking full account of the difference in basis sets. 
The materials can have very different bulk electronic structures, which are properly handled, equivalent to the Foreman technique.
However, we do so by expanding on a single basis set, which leads to a continuous set of equations free of interface terms, which seems appropriate given a single Hilbert space for the solution to the entire heterostructure.
Neither new parameters, nor additional assumptions regarding the material properties are in order,
only the standard $\vk\cdot\vp$ matrix elements of every material involved need to be known.
A benchmark using this new formalism to calculate band-to-band transitions (BTBT) in heterostructures has been presented and published\cite{VandePut:2013up}, and more recently, a two dimensional BTBT simulator has been implemented\cite{Verreck:2014di}.
In the present paper, we elaborate on the underlying theory and its implications.

\section{Heterostructure model: The Schrödinger equation}

The goal of this paper is to develop an envelope function formalism to solve the one-electron Schrödinger equation in lattice-matched semiconductor heterostructures under the Hartree approximation,
\begin{equation}
  -\frac{\hbar^2}{2m_0}\vnabla^2 \psi(\vr) + V_{\rm c}(\vr)\,\psi(\vr) + V_{\rm e}(\vr)\,\psi(\vr) = E\,\psi(\vr),
  \label{schroedinger}
\end{equation}
where the crystal potential $V_{\rm c}(\vr)$ is taken piecewise throughout the structure and the extrinsic potential $V_{\rm e}(\vr)$ contains all contributions to the potential energy not related to the bulk material.

\begin{figure}[h!]
  \begin{center}
  \includegraphics[width=0.4\linewidth]{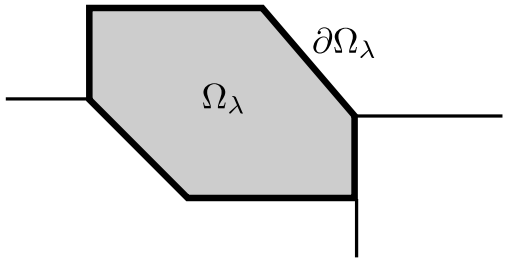}
  \end{center}
  \caption{\label{heterostructure} Part of a general heterostructure with a region $\Omega_l$ and its boundary $\partial\Omega_l$ shown.}
\end{figure}

We distinguish between the heterostructure materials with an index $\lambda$, each material having a region $\Omega_\lambda$ in the heterostructure and a boundary $\partial\Omega_\lambda$ as depicted in Fig.~\ref{heterostructure}.

Using this notation we can write the full form of the crystal potential,
\begin{equation}
  V_{\rm c}(\vr) = \sum\limits_\lambda \theta(\vr \in \Omega_\lambda) V^\lambda_{\rm c}(\vr),
  \label{crystal_potential}
\end{equation}
where the sum runs over all materials $\lambda$. $\theta(\vr\in\Omega_\lambda)$ is the logical step function, or indicator function, which equals $1$ if $\vr$ is in $\Omega_\lambda$, $0$ outside this region and it selects the correct crystal potential for a given region $\lambda$. The crystal potentials $V_{\rm c}^\lambda(\vr)$ are lattice periodic and defined in all space.

\section{An arbitrary basis}
We expand the wavefunction in an arbitrary set of complete, periodic basis functions $u_n(\vr)$, yielding
\begin{equation}
  \psi(\vr) = \sum\limits_n f_n(\vr) u_n(\vr),
  \label{e:expansionwf}
\end{equation}
where the expansion coefficients $f_n(\vr)$ are called envelope functions. Due to the periodicity of the basis functions, this expansion is unique if the plane wave expansion of the envelope functions only has non-zero components within the first Brillouin zone.

Substitution into the Schrödinger equation Eq.~(\ref{schroedinger}) leads to
\begin{multline}
  \sum\limits_n \Bigg[ 
    -\frac{\hbar^2}{2m_0} \vnabla^2 f_n(\vr) u_n(\vr)
    - \frac{\im\hbar}{m_0} \vnabla f_n(\vr) \cdot (-\im\hbar\vnabla) u_n(\vr) 
    + f_n(\vr) \frac{-\hbar^2}{2m_0} \vnabla^2 u_n(\vr) \\
    + V_{\rm c}(\vr) f_n(\vr) u_n(\vr) 
    + V_{\rm e}(\vr) f_n(\vr) u_n(\vr)
  \Bigg]
  = \sum\limits_n E\, f_n(\vr) u_n(\vr).
  \label{full_expanded_envelope}
\end{multline}
In the next sections we introduce momentum and energy matrix elements in order to recast Eq.~(\ref{full_expanded_envelope}) into a set of coupled equations in which the differential operators merely act on the envelope functions.

\subsection{\label{ss:momentum}Momentum term}
Starting with the second term of Eq.~(\ref{full_expanded_envelope}), we write $-\im\hbar\vnabla u_n(\vr)$ as an integral over a unit cell,
\begin{equation}
  -\im\hbar\vnabla u_n(\vr) = \int\limits_{\rm uc} \rd^3r' \delta(\vr-\vr') (-\im\hbar\vnabla') u_n(\vr'),
\end{equation}
and exploit the completeness relation $\sum_m u^*_m(\vr') u_m(\vr) = \delta(\vr-\vr')$, yielding
\begin{equation}
  -\im\hbar\vnabla u_n(\vr) = \sum\limits_m \int\limits_{\rm uc} \rd^3r' 
      u_m(\vr') (-\im\hbar\vnabla') u_n(\vr') u_m(\vr).
\end{equation}
where the periodicity of the basis functions restricts the integration to a single unit cell (uc).
Substitution into the second term of Eq.~(\ref{full_expanded_envelope}) and swapping $n$ and $m$ leads to
\begin{equation}
  \sum\limits_{nm} -\frac{\im\hbar}{m_0} \vnabla f_m(\vr) \cdot \vp_{nm} u_n(\vr).
\end{equation}
The momentum matrix elements are empirically determined for basis functions which are solutions for a bulk material corresponding to a point of high symmetry in the first Brillouin zone,
\begin{equation}
  \vp_{nm} = -\im\hbar \int\limits_{\rm uc} {\rm d}^3r\, u^*_n(\vr) \vnabla u_m(\vr).
  \label{arb_me_mom}
\end{equation}

\subsection{\label{ss:energy}Bulk Hamiltonian term}

Combining the third an fourth term of Eq.(\ref{full_expanded_envelope}), we recover the bulk Hamiltonian at $\v{k}=(000)$ of layer $\lambda$,
\begin{equation}
  \sum\limits_n \left[ -\frac{\hbar^2}{2m_0} \vnabla^2 + V_{\rm c}(\vr) \right] u_n(\vr) f_n(\vr)
\end{equation}
Expanding the crystal potential as in Eq.~(\ref{crystal_potential}) and using the fact that $\sum_\lambda\theta(\vr\in\Omega_\lambda) = 1$ we obtain
\begin{equation}
  \sum\limits_n \sum\limits_\lambda \theta(\vr\in\Omega_\lambda) 
    \left[ -\frac{\hbar^2}{2m_0} \vnabla^2 + V^\lambda_{\rm c}(\vr) \right] u_n(\vr) f_n(\vr)
\end{equation}
Inserting a completeness relation for the two terms and swapping indices $n$ and $m$, we obtain an expression in the basis functions directly,
\begin{equation}
  \sum\limits_{nm} \sum\limits_\lambda \theta(\vr\in\Omega_\lambda) H^\lambda_{nm} f_m(\vr)
  u_n(\vr),
  \label{e:hujhdfscaglijkhsadrhgliu}
\end{equation}
where the matrix elements of the bulk Hamiltonian for layer $\lambda$ are given by
\begin{equation}
  H^\lambda_{nm} = \int\limits_{\rm uc} \rd^3r\, u^*_n(\vr) \left[ -\frac{\hbar^2}{2m_0}\vnabla^2 + V^\lambda_{\rm c}(\vr) \right] u_m(\vr).
  \label{arb_me_bh}
\end{equation}

\subsection{\label{ss:ef_equation}Envelope function equation}
Using the completeness of the $u_n(\vr)$, we arrive at a set of envelope function equations
\begin{equation}
  - \frac{\hbar^2}{2m_0} \vnabla^2 f_n(\vr)
  - \frac{\im\hbar}{m_0} \sum\limits_m \vp_{nm} \cdot \vnabla f_m(\vr) 
  + \sum\limits_m H_{nm}(\vr) f_m(\vr) 
  + V_{\rm e}(\vr) f_n(\vr) 
  = E\, f_n(\vr),
  \label{undef_envelope_function_equation}
\end{equation}
with momentum matrix elements $\vp_{nm}$ as defined in Eq.~(\ref{arb_me_mom}), and the position dependent bulk-Hamiltonian matrix elements
\begin{equation}
  H_{nm}(\vr) = \sum\limits_\lambda \theta(\vr\in\Omega_\lambda) H^\lambda_{nm},
  \label{arb_me_bh_pd}
\end{equation}
which use the material dependent bulk Hamiltonian matrix elements $H^\lambda_{nm}$ as in Eq.~(\ref{arb_me_bh}).

Because the envelope functions do not have to be restricted to the first Brillouin zone, the expansion of the wavefunctions as specified in Eq.~(\ref{e:expansionwf}) is not unique, but rather convenient. A derivation of envelope functions which are unique, but non-local by construction, is detailed in the appendix.

\section{\label{s:problem} Selecting a basis set: the heterostructure problem}
As mentioned before, we need to select a specific set of basis functions such that the matrix elements can be empirically determined. 

Choosing the ${u_n^\rho(\vr)}$ basis to be the zone-centre solutions of the bulk material denoted by index $\rho$ we may write the envelope function expansion of the wavefunction as
\begin{equation}
  \psi(\vr) = \sum\limits_n f^\rho_n(\vr) u^\rho_n(\vr),
\end{equation}
whereas the basis dependent envelope functions $f^\rho_n(\vr)$ satisfy the envelope function equations given by
\begin{equation}
  - \frac{\hbar^2}{2m_0} \vnabla^2 f^\rho_n(\vr)
  - \frac{\im\hbar}{m_0} \sum\limits_m \vp^\rho_{nm} \cdot \vnabla f^\rho_m(\vr) 
  + \sum\limits_m H^\rho_{nm}(\vr) f^\rho_m(\vr) 
  + V_{\rm e}(\vr) f^\rho_n(\vr) 
  = E\, f^\rho_n(\vr).
  \label{envelope_function_equation}
\end{equation}
Here, the inter-band momentum matrix elements $\vp^\rho_{nm}$ only depend on the chosen basis and are known for many basis sets.\cite{Vurgaftman:2001bu} Being independent on the local material, they are position independent.

The bulk Hamiltonian matrix elements are dependent on the basis through $\rho$, and the local crystal potential through $\lambda$,
\begin{equation}
H_{nm}^{\rho,\lambda} = \int \rd^3r\, u^{\rho*}_n(\vr) 
    \left[ -\frac{\hbar^2}{2m_0}\vnabla^2 + V^\lambda_{\rm c}(\vr) \right] u^\rho_m(\vr).
\label{matdep_bulk_ham}
\end{equation}
They are well known when $\rho=\lambda$, namely
\begin{equation}
H^{\lambda,\lambda}_{nm} = \delta_{n,m} E^\lambda_n,
\end{equation}
where $E^\lambda_n$ are the band zone-centre energies associated with the Bloch basis $\{u_n^\lambda(\vr)\}$.

However, when considering $H_{nm}^{\rho,\lambda}$ with $\rho\neq\lambda$ we cannot rely on empirically obtained data and one would need to evaluate the integral in Eq.~(\ref{matdep_bulk_ham}), as this would require full knowledge of the zone-centre basis set $\{u_n^\rho(\vr)\}$ and the mean-field crystal potentials $V^\lambda_{\rm c}(\vr)$ for all $\lambda$. Moreover the integral would have to be computed numerically in general.

\section{\label{s:transformation}Basis function transformation}

To circumvent the problem of the unknown matrix elements we perform a unitary transformation $S^{\lambda\to\rho}$ which transforms the ${u_m^\lambda(\vr)}$ basis into ${u_n^\rho(\vr)}$, i.e.
\begin{equation}
  u^\rho_n(\vr) = \sum\limits_m S_{nm}^{\lambda\to\rho}\, u^\lambda_m(\vr),
\end{equation}
where
\begin{equation}
  S_{nm}^{\lambda\to\rho} = \int {\rm d}^3r \, u^{\rho*}_n(\vr) u^\lambda_m(\vr).
  \label{transform_element_full}
\end{equation}

We could, in principle at least, determine the transformation matrix elements from the set of basis functions $\{u^\rho_{nm}(\vr)\}$ by numerical integration of Eq.~(\ref{transform_element_full}). However, a central point to the practical use of $\vk\cdot \vp$ theory is that we don't need to know the basis functions. Finding the basis function requires the solution of the full bulk Schr\"odinger equation at the $\vk=(000)$ point, which in turn requires knowledge of the crystal potential $V_{\rm c}^\lambda(\vr) \forall \lambda$. With $\vk\cdot\vp$, we avoid this problem by introducing empirically determined inter-band momentum matrix elements $p_{nm}$ and zone-centre energies $E_n$. In this section we will detail a method in which the elements of the transformation matrix $S_{nm}^{\lambda\to\rho}$ are be obtained solely from these empirical parameters.

\subsection{\label{ss:selems}The transformation matrix elements}

We set out by introducing a basis set of plane wave with lattice periodicity, namely 
\begin{equation}
  u_\vG(\vr) = e^{i\vG\cdot\vr},
\end{equation}
where $\vG$ is a reciprocal lattice vector.

As the plane wave basis functions are eigenfunctions of the momentum operator the inter-band momentum matrix elements turn out to be diagonal,
\begin{equation}
  \vp_{\vG\vG'} 
  = -\im\hbar \int {\rm d}^3r\, u^*_\vG(\vr) \vnabla u_\vG(\vr) 
  = \delta_{\vG\vG'}\,\hbar\vG.
  \label{e:G_mom_mat}
\end{equation}

Eigenvalue decomposition of the inter-band momentum matrix $\v{P}^\lambda$ of any complete set of basis functions ${u_n^\lambda(\vr)}$ with the same lattice periodicity will produce these exact same eigenvalues:
\begin{equation}
  \v{P}^\lambda = R^\lambda\ {\rm diag}(\hbar\vG) R^{\lambda\dagger},
  \label{e:p_eig_decomp}
\end{equation}
where we assume the eigenvectors in the columns of $R^\lambda$ are properly normalized so that $R^\lambda$ is a unitary matrix. 
The momentum matrix elements read
\begin{equation}
  \vp_{nm}^\lambda 
  = \sum\limits_\vG R_{n\vG}^\lambda\, \hbar\vG\, R_{\vG m}^{\lambda\dagger}
  = \sum\limits_\vG R_{n\vG}^\lambda\, \hbar\vG\, R_{m\vG}^{\lambda*}
  \label{e:eigendecomp_components}
\end{equation}
where the matrix elements $R_{n\vG}^\lambda$ are the eigenvectors of the inter-band momentum matrix $\v{P}^\lambda$. 

By defining a transformation $Q^\lambda$ from the plane wave basis ${u_\vG(\vr)}$ to the basis functions ${u^\lambda_n(\vr)}$ of material $\lambda$,
\begin{equation}
  u^\lambda_n(\vr) = \sum\limits_{\vG} Q^\lambda_{nG} u_\vG(\vr),
\end{equation}
and using this transformation on the inter-band momentum matrix, we obtain a similar equation as in Eq.~(\ref{e:eigendecomp_components}),
\begin{align}
  \vp^\lambda_{nm} 
  &= -\im\hbar \int {\rm d}^3r\, u^{\lambda*}_n(\vr) \vnabla u^\lambda_m(\vr) \nonumber\\
  &= -\im\hbar \sum\limits_{\vG\vG'} \int {\rm d}^3r\, 
    Q^{\lambda*}_{n\vG} u^*_\vG(\vr) \vnabla Q^{\lambda}_{m\vG} u_{\vG'}(\vr) \nonumber\\
  &= \sum\limits_\vG Q^{\lambda*}_{n\vG}\,\hbar\vG\,Q^{\lambda}_{m\vG},
\end{align}
or in matrix-representation
\begin{equation}
  \v{P}^\lambda = Q^{\lambda*}\,\hbar\vG\,Q^{\lambda\intercal},
\end{equation}
from which we conclude $R^\lambda=Q^{\lambda*}$. The matrix elements $Q^\lambda_{n\vG}$ of the transformation from plane waves to the basis set of material $\lambda$ are completely determined by the eigendecomposition of the inter-band momentum matrix of the corresponding material. The inter-band momentum matrix elements can be obtained from $\vk\cdot\vp$ models with a high number of bands, readily available in literature for common materials.\cite{Richard:2004wo,Fraj:2007wr}
  
For the transformation between two material basis sets ${u^\lambda_n(\vr)}$ and ${u^\rho_n(\vr)}$, we derive
\begin{align}
  S_{nm}^{\lambda\to\rho} 
  &= \int \rd^3r\, u_n^{\rho*}(\vr) u_m^\lambda(\vr) 
  = \int \rd^3r \sum\limits_\vG \sum\limits_{\vG'} 
    Q_{n\vG}^{\rho*} e^{-\im\vG\cdot\vr} Q_{m\vG'}^\lambda e^{\im\vG\cdot\vr} \nonumber\\
  &= \sum\limits_\vG \sum\limits_{\vG'} Q_{n\vG}^{\rho*}\, Q_{m\vG'}^\lambda \delta_{\vG,\vG'} 
  = \sum\limits_\vG Q_{n\vG}^{\rho*}\, Q_{m\vG}^\lambda
\end{align}
or in matrix format:
\begin{equation}
  S^{\lambda\to\rho} = Q^{\rho\dagger}\, Q^{\lambda}.
  \label{e:subs_transf}
\end{equation}

In order to obtain the transformation matrix $S^{\lambda\to\rho}$ we do not need to evaluate the integral in Eq.~(\ref{transform_element_full}). Instead, we need only to diagonalize the inter-band momentum matrix $\v{P}^\lambda$ so as to obtain $Q^\lambda$, while the inter-band momentum matrix is a well known object. 

\subsection{\label{ss:t_elems}Revisiting the bulk Hamiltonian matrix elements}

With the transformation matrix elements known, we can now easily express the transformation matrix elements $H^{\lambda,\rho}_{nm}$ of Eq.~\ref{envelope_function_equation} in terms of the available material specific matrix elements by transforming from the reference basis set to the zone-centre solutions $u^\lambda_n(\vr)$,
\begin{align}
  H^{\lambda,\rho}_{nm}
    &= \int {\rm d}^3r\, 
      u^{\rho*}_n(\vr) \left[-\frac{\hbar^2}{2m} \vnabla^2 + V_{\rm c}^\lambda(\vr) \right] u^\rho_m(\vr)\nonumber\\
    &= \sum\limits_{ij} \int {\rm d}^3r\, 
      u^{\lambda*}_i(\vr) \left[S^{\lambda\to\rho}_{ni}\right]^*
      \hat{H}^\lambda S^{\lambda\to\rho}_{mj} u^\lambda_j(\vr)\nonumber\\
    &= \sum\limits_{ij} \left[S^{\lambda\to\rho}\right]^*_{ni} H^{\lambda,\lambda}_{ij} S^{\lambda\to\rho}_{mj}
\end{align}
Because the bulk Hamiltonian of material $\lambda$ is now expressed in it's own natural basis set the matrix elements reduce to the band energies at the zone-centre $H^{\lambda,\lambda}_{ij} = \delta_{ij} E^\lambda_i$. The bulk-Hamiltonian matrix elements of material $\lambda$ in the ${u^\rho_n(\vr)}$ basis are thus found as
\begin{equation}
  H^{\lambda,\rho}_{nm} = \sum\limits_i \left[S^{\lambda\to\rho}_{ni}\right]^* E^\lambda_i S^{\lambda\to\rho}_{mi}
  \label{e:hm_transf}
\end{equation}
As expected, the columns of the complex conjugated transformation matrix are the eigenvectors of the bulk-Hamiltonian of a material $\lambda$ expressed in the reference basis ${u_n^\rho(\vr)}$. This is equivalent to the momentum matrix decomposition that led to the transformation matrix elements.

\subsection{Transforming the envelope functions}

The envelope functions are expansion coefficients with respect to the chosen reference basis set. In most applications, it is however convenient to have envelope functions defined against the material's own zone-centre bulk-Hamiltonian solutions. We achieve this by transforming the calculated envelope functions in each material to it's proper material specific basis set.

In the region of material $\lambda$ we transform to the basis functions $u^\lambda_n(\vr)$,
\begin{equation}
  \psi(\vr) = \sum\limits_m f^\rho_m(\vr) u^\rho_m(\vr) 
            = \sum\limits_{mn} f^\rho_m(\vr) S^{\lambda\to\rho}_{mn} u^\lambda_n(\vr)\nonumber
            = \sum\limits_n f^\lambda_n(\vr) u^\lambda_n(\vr),
\end{equation}
which amounts to
\begin{equation}
  f^\lambda_n(\vr) = \sum\limits_{m} f^\rho_m(\vr) S^{\lambda\to\rho}_{mn}. 
\end{equation}
These envelope functions can be interpreted just like those produced by the homostructure envelope function formalism.

\section{\label{s:basis_choice}Basis choice and restriction}

While the basis transformation enables envelope function expansion with respect to a particular reference basis set, one is free to select that reference basis set.
In this section, we investigate the effects of this choice when the set of basis functions is truncated such that the number of basis functions is finite. 
The truncated basis set is no longer complete and does not span the whole solution space of the original Schr\"odinger equation.
When changing basis sets, we have two options for the momentum matrix elements: either the momentum matrix elements are transformed like the Hamiltonian matrix elements, i.e. ${\bf P}^{\lambda}\to\left[S^{\lambda\to\rho}\right]^{\dagger}{\bf P}^{\lambda}S^{\lambda\to\rho}$, or the momentum matrix elements ${\bf P}^{\rho}$ from the new basis are used.
We show that the former option leads to a set of equations that is invariant under basis choice.

\subsection{\label{s:2-band}A two band model}
\newcommand{\fvtwo}{\left[\begin{array}{c}f^\rho_1(x)\\f^\rho_2(x)\end{array}\right]}
\newcommand{\fvtwol}{\left[\begin{array}{c}f^\lambda_1(x)\\f^\lambda_2(x)\end{array}\right]}

In this section, we study the simplest case by restricting the basis set to just two functions $u^\rho_1(\vr)$ and $u^\rho_2(\vr)$, exhibiting a non-zero inter-band momentum matrix element between them only in the $x$ direction,
\begin{equation}
  \vp^\rho_{12} = -\im\hbar \int \rd^3r\, u_1^{\rho*}(\vr) \nabla u^\rho_2(\vr) = (p^\rho,0,0).
\end{equation}
One can think of the conduction band state with $\rm s$-like symmetry and a valence band state with $\rm p_x$-like symmetry in the $\Gamma$-point of semiconductors. In this simple model a major problem arises due to the freedom of choice for the reference basis set. This problem will be resolved in the next section

The coupling being one-dimensional, Eq.~(\ref{envelope_function_equation}) now reduces to
\begin{equation}
  - \frac{\hbar^2}{2m_0} \dd{}{x} \fvtwo
  - \frac{\im\hbar}{m_0} P^\rho \d{}{x} \fvtwo
  + H^\rho(x) \fvtwo
  + V_{\rm e}(x) \fvtwo
  = E\, \fvtwo,
\end{equation}
where the envelope function are $f_1^\rho(x)$ and $f_2^\rho(x)$, with a simple inter-band momentum matrix,
\begin{equation}
P^\rho = \left[\begin{array}{cc} 0 & p^\rho\, \\ \,p^{\rho*} & 0\, \end{array}\right].
\end{equation}

In this oversimplified case it is not even necessary to calculate the transformation matrix elements as the symmetry considerations lead exclusively to the identity transformation,
\begin{equation}
S^{\rho\to\lambda} = \left[\begin{array}{cc} \,1 & 0\, \\ \,0 & 1\, \end{array}\right],
\end{equation}
for every lattice matched material combination $\lambda$, $\rho$. 

In the two-band case, when using our basis transformation model, there is no distinction between the basis functions of the different materials and this results in a very simple set of envelope function equations. Only the basis function energies $E_1$ and $E_2$ are material dependent,
\begin{equation}
  \left\{- \frac{\hbar^2}{2m_0} \dd{}{x}
  - \frac{\im\hbar}{m_0} 
      \left[\begin{array}{cc}0 & p^\rho\\(p^\rho)^* & 0\end{array}\right] \d{}{x} 
  + \sum\limits_\lambda \theta(x\in\Omega_\lambda) 
      \left[\begin{array}{cc} E_{\rm 1}^\lambda & 0 \\ 0 & E_{\rm 2}^\lambda \end{array}\right]
  + V_{\rm e}(x) \right\} \fvtwo
  = E\, \fvtwo.
  \label{e:twoband_wrong}
\end{equation}
The inability of the transformation to distinguish the basis functions of two different materials and the envelope function equations defined earlier has introduced an ambiguity. Changing the basis changes the inter-band momentum matrix elements but not the band edge energies and results in different equations with different solutions.

To make the ambiguity explicit, we consider a bulk material $\lambda$, without any external potential. The envelope function equations in the basis set $u^\lambda_n(x)$ yield the expected $\vk\cdot\vp$ equations,
\begin{equation}
  \left\{- \frac{\hbar^2}{2m_0} \dd{}{x} 
  - \frac{\im\hbar}{m_0} 
      \left[\begin{array}{cc}0 & p^\lambda\\(p^\lambda)^* & 0\end{array}\right] \d{}{x}
  + \left[\begin{array}{cc} E_{\rm c}^\lambda & 0 \\ 0 & E_{\rm v}^\lambda \end{array} \right]
  + V_{\rm e}(x) \right\} \fvtwol
  = E\, \fvtwol
  \label{e:2band_1}
\end{equation}
but transforming the into a different basis set ${u_n^\rho(x)}$, Eq.~(\ref{e:2band_1}) becomes
\begin{equation}
  \left\{- \frac{\hbar^2}{2m_0} \dd{}{x}
  - \frac{\im\hbar}{m_0} 
      \left[\begin{array}{cc}0 & p^\rho\\(p^\rho)^* & 0\end{array}\right] \d{}{x}
  + \left[\begin{array}{cc} E_{\rm c}^\lambda & 0 \\ 0 & E_{\rm v}^\lambda \end{array}\right]
  + V_{\rm e}(x) \right\} \fvtwo
  = E\, \fvtwo
\end{equation}
and in general $p^\lambda \neq p^\rho$, so our model seems to provide the wrong $\vk\cdot\vp$ equation for bulk material $\lambda$ when expressed in any basis set ${u_n^\lambda(x)}$ other than its own. 

Hence, to remedy this problem, we should use a material dependent inter-band momentum matrix element, giving rise to the following two band envelope function equations:
\begin{multline}
  \left\{- \frac{\hbar^2}{2m_0} \dd{}{x}
  - \frac{i\hbar}{m_0} \sum\limits_\lambda \theta(x\in\Omega_\lambda)
    \left[\begin{array}{cc}0 & p^\lambda\\(p^\lambda)^* & 0\end{array}\right] \frac{-\im\hbar}{m_0} \d{}{x}
  + \sum\limits_\lambda \theta(x\in\Omega_\lambda)
    \left[\begin{array}{cc} E_{\rm c}^\lambda & 0 \\ 0 & E_{\rm v}^\lambda \end{array}\right] 
     \vphantom{\left[\begin{array}{cc} E_{\rm c}^\lambda & 0 \\ 0 & E_{\rm v}^\lambda \end{array}\right]}
  + V_{\rm e}(x) \right\} \fvtwo \\
  = E\, \fvtwo.
\end{multline}
One can easily check that these equations do produce the right $\vk\cdot\vp$ equations for bulk material, for any basis.

In the two band case, our heterostructure formalism with basis function transformations is equivalent to taking position dependent matrix elements in the classical envelope function method without any transformations.

\subsection{Robust envelope function equations}
The above considerations clearly indicate that using a finite number of basis functions calls for transformed momentum matrix elements in the envelope function equations. In the infinite complete basis set we have
\begin{equation}
 \vp^\rho_{nm} = \sum\limits_{ij} \left[S^{\lambda\to\rho}_{ni}\right]^* \vp^\lambda_{ij} S^{\lambda\to\rho}_{mj}.
\end{equation}
with which we obtain a modified set of envelope function equations,
\begin{equation}
  - \frac{\hbar^2}{2m_0} \vnabla^2 f^\rho_n(\vr)
  - \frac{\im\hbar}{m_0} \sum\limits_m \vp^\rho_{nm}(\vr) \cdot \vnabla f^\rho_m(\vr) 
  + \sum\limits_m H^\rho_{nm}(\vr) f^\rho_m(\vr) 
  + V_{\rm e}(\vr) f^\rho_n(\vr)
  = E\, f^\rho_n(\vr),
  \label{robust_envelope_function_equation}
\end{equation}
with position dependent momentum matrix elements,
\begin{align}
  \vp^\rho_{nm}(\vr) &= \sum\limits_\lambda \theta(\vr\in\Omega_\lambda) \,\vp^{\lambda,\rho}_{nm},\\
  \vp^{\lambda,\rho}_{nm} &= \sum\limits_{ij} 
     \left[S^{\lambda\to\rho}_{in}\right]^* p^\lambda_{ij} S^{\lambda\to\rho}_{jm}.
\end{align}
While, in the case of a complete basis set, the exact relation $\vp^{\lambda,\rho}_{nm} = \vp^\rho_{nm}$ still holds, using this new, robust form of envelope equation the choice of reference basis set has no influence on the final results, even when these basis sets are as heavily restricted as in the two band model.
This can be verified for any arbitrary restriction by performing a transformation of the envelope functions $f_n^\rho(\vr)$ to the $f_m^\rho(\vr)$ basis set which results in pairs of transformation matrices $[S^{\lambda\to\rho}]^\dagger S^{\lambda\to\rho}$ that cancel because of unitarity, even for an incomplete basis set.

Furthermore, because of the symmetry constraints imposed on the basis set, this form results in the same equations used in literature whenever the set is restricted to bands with all different symmetry as is the case in the two band model, but also in the four (eight) band model of a semiconductor with conduction band states with $\rm s$-like symmetry and valence states  with $\rm p_x$-, $\rm p_y$- and $\rm p_z$-like symmetry. However, for more accurate models such as those including 15 or 30 bands\cite{Richard:2004wo,Richard:2005wl,saidi2010band}, this form will provide more accurate results, certainly when inter-band processes are important.

\section{\label{s:observables}Unit-cell observables}

Being represented by operators acting on the entire wavefunction, observables are independent of the chosen basis.
It is however desirable to define operators acting only on the envelope function. These can only give approximate results for the observables because they are defined by averaging over one unit-cell to remove the basis function dependence.
In this section, we show expressions for the unit-cell average density and probability current.
We show that these observables can be determined from the envelope functions in any basis set without explicit transformation to the original material basis.

We define the average unit-cell density of a pure state $\psi(\vr)$ as
\begin{equation}
\left<\rho\right>_{{\rm uc}}(\vr)=\sum\limits _{n}f_{n}^{\rho*}(\vr)f_{n}^{\rho}(\vr)
\end{equation}
and since any transformation to another basis set $f_{n}^{\rho}(\vr)$ is unitary, the average unit-cell density is basis independent.

Next, we turn to the probability current. The probability current is given by
\begin{equation}
  \v{J}_{\rm prob}(\vr) = -\frac{\im\hbar}{2m_0} 
    \left[ \psi^*(\vr) \vnabla \psi(\vr) - \psi(\vr) \vnabla \psi^*(\vr) \right].
\end{equation}
We use the envelope function expansion and take an average over one unit cell, removing the basis function dependence and introducing inter-band momentum matrix elements in the equation,
\begin{equation}
  \left<\v{J}_{\rm prob}\right>_{\rm uc}(\vr) =
    - \frac{\im\hbar}{2m_0} \sum\limits_n 
      \left[ f^{\rho*}_n(\vr) \vnabla f^\rho_n(\vr) - f^\rho_n(\vr) \vnabla f^{\rho*}_n(\vr) \right] 
    + \frac{1}{m_0} \sum\limits_{n,m} \vp^\rho_{nm}(r) f^{\rho*}_n(\vr) f^\rho_m(\vr).
    \label{uc.avg.prob.curr}
\end{equation}
Here we have again used the position dependent matrix element to make Eq.~(\ref{uc.avg.prob.curr}) robust to basis changes when considering an incomplete basis set. Upon transformation of the envelope functions to another basis we find that it is basis independent due to unitarity of the transformation.

\section{\label{s:conclusions}Conclusions}
We have proposed a transformation between zone-centre solutions of the bulk Hamiltonian of different materials. The transformation is fully described by the eigendecomposition of the inter-band momentum matrices of both materials. The inter-band momentum matrices are empirically determined by $\vk\cdot\vp$ fitting to bulk material and available in literature for many materials.

This transformation enables the expansion of the wave function on the zone-centre solutions of any particular material. 
Without this transformation we are confronted with unknown matrix elements that cannot be empirically determined. 
In our approach, we determined the unknown matrix elements by transforming them from a natural basis set where they are known to a singular reference basis set which is used as a basis for the envelope expansion. No new parameters were needed in this process.

Our method performs a similar function to previously published work, and differs from it mostly in a practical sense. For example the original formalism of Burt\cite{Burt:1992,Burt:1999cw} allows for heterostructures and takes account of non-local effects due to abrupt changes in potential. However, in application of this method it is assumed that the basis functions are known. As stated earlier this is not the case for empirically fitted $\vk\cdot\vp$ models. Our method is also closely related to the basis-dependent method by Foreman\cite{Foreman:1996gt} as used in \cite{morandi:2014}, where the basis functions change with the material while interface terms mediate the coupling (transformation) from one material to the next. Our method effectively uses a single basis set throughout the whole structure and instead transforms the $\vk\cdot\vp$ parameters in each region to this shared basis. Because we expand on a single basis set no explicit interface terms appear.

We should note that the non-local caused by abrupt changes of the potential present in the formalisms by Burt and Foreman are also applicable to our method and can be found in the Appendix. This also includes non-local effects caused by abrupt changes in momentum and energetic coupling between materials.

We have taken care to make the heterostructure envelope functions robust under change of basis when this basis is no longer complete. This method is thus applicable even for only a few bands with the same accuracy as the homostructure based approach. For the basis set of two coupled bands, and other basis sets with heavy symmetry restrictions, we obtained a simpler set of equations, where in each region the homostructure equation is retrieved and no transformation is present.

To conclude, we note that the computational burden related to this method is no greater than for the conventional homostructure methods. The transformation matrix calculation involves a computationally inexpensive simultaneous eigenvalue decomposition of the 3D inter-band momentum matrix. The transformation matrix can even be tabulated in advance of any calculations.

\appendix
\section*{Appendix: Unique envelope functions}
\setcounter{section}{1}
\label{a:uniqueenvelope}
The heterostructure envelope function equations discussed in this paper do not correspond to a unique expansion of the wavefunction. As we show in this section, a unique expansion results in non-local integro-differential envelope function equations, which are not as convenient. The convenient, local, if not unique, equations discussed throughout the paper are based on the slowly-varying field approximation commonly used in practical envelope function calculations. This means that the external potential and the matrix elements must not change abruptly on the scale of a unit cell. For the external potential, this should be evaluated on a case-by-case basis. As for the matrix elements, we can appeal to physical demands that the basis set and crystal potential do not contain abrupt discontinuities due to the non-local nature of the electron interaction with the lattice.

It is however possible to derive envelope function equations for a unique expansion of the basis set. We enforce this uniqueness by restricting the non-zero plane wave components of the envelope functions to the first Brillouin Zone (1BZ). 
 We obtain envelope function equations that return exact and unique envelope functions at the cost of locality of the envelope function equations.

\subsection{External potential}
For the external potential, we adopt Burt's methodology to restrict the plane wave components to the 1BZ, yielding\cite{Burt:1988fa}
\begin{equation}
\sum\limits_{n} V_{\rm e}(\vr) f_n(\vr) u_n(\vr) 
= \sum\limits_{nm} \int \rd^3r' V_{nm}(\vr,\vr') f_m(\vr') u_n(\vr),
\label{e:non_local_V}
\end{equation}
where the non-local kernel $V_{nm}(\vr,\vr')$ is
\begin{equation}
V_{nm}(\vr,\vr') = \sum\limits_{\vk\vk'} \sum\limits_{\vG\vG'} 
\left[ \tilde{u}_{n\vG+\vG_1} \right]^* \tilde{V}_{\vG-\vG'}(\vk) \tilde{u}_{m\vG'} 
e^{\im(\vk_1\cdot\vr-\vk'\cdot\vr')}.
\end{equation}
Here, $\tilde{u}_{n\vG}$ and $\tilde{V}_\vG(\vk)$ are the Fourier transform of $u_n(\vr)$ and $V_{\rm e}(\vr)$ respectively. $\vk_1$ is a wavevector inside the 1BZ and $\vG_1$ is a reciprocal lattice vector defined by $\vk+\vk'=\vk_1+\vG_1$.

Invoking a basis set transformation, we have now also determined the Fourier components by the eigendecomposition of the momentum matrix as $\tilde{u}^\rho_{n\vG} = Q_{n\vG}$. The exact knowledge of the basis functions is thus no longer needed to determine the non-local interaction.

\subsection{Bulk Hamiltonian matrix elements}
We apply a similar procedure to the bulk Hamiltonian matrix elements, starting with the expansion of Eq.~(\ref{e:hujhdfscaglijkhsadrhgliu}) into plane waves,
\begin{equation}
\sum\limits_{nm} H_{nm}(\vr) f_m(\vr) u_n(\vr)
= \sum\limits_{nm}\sum\limits_{\vk\vk'}\sum\limits_{\vG\vG'} \tilde{H}_{nm,\vG}(k) \tilde{f}_m(\vk') \tilde{u}_{n\vG'} e^{\im(\vk+\vG+\vk'+\vG')\cdot\vr}.
\end{equation}
Substituting $\vG \rightarrow \vG-\vG'$, expressing $\vk+\vk'$ as a 1BZ restricted wavevector $\vk_1$, and introducing a reciprocal lattice vector $\vG_1$ with $\vk+\vk=\vk_1+\vG_1$,
\begin{equation}
\sum\limits_{nm}\sum\limits_{\vk\vk'}\sum\limits_{GG'} \tilde{H}_{nm,\vG-\vG'}(k) \tilde{f}_m(\vk') \tilde{u}_{n\vG'} e^{\im\vk_1\cdot\vr} e^{\im(\vG+\vG_1)\cdot\vr}
\end{equation}
we finally rewrite $e^{\im(\vG+\vG_1)\cdot\vr}$ using the basis functions $e^{\im(\vG+\vG_1)\cdot\vr} = \sum\limits_j \tilde{u}_{j,\vG+\vG_1}^* u_j(\vr)$,
\begin{equation}
\sum\limits_{nmj}\sum\limits_{\vk\vk'}\sum\limits_{\vG\vG'} \tilde{H}_{nm,\vG-\vG'}(\vk) \tilde{f}_m(\vk') \tilde{u}_{n\vG'}  \tilde{u}_{j,\vG+\vG_1}^* e^{\im\vk_1\cdot\vr} u_j(\vr).
\end{equation}
We obtained an expansion on the basis functions ${u_j(\vr)}$ where the expansion coefficients are restricted to the 1BZ because $\vk_1$ is always inside the 1BZ. Performing a inverse Fourier decomposition on the envelope functions $\tilde{f}_m(\vk') = \int \rd^3r' f_m(\vr) e^{-\im\vk'\cdot\vr'}$ and exchanging indices $n$ and $j$ yields
\begin{equation}
\sum\limits_{nm} H_{nm}(\vr) f_m(\vr) u_n(\vr)
= \sum\limits_n \left[ \sum\limits_m \int \rd^3r' H_{nm}(\vr,\vr') f_m(\vr') \right] u_n(\vr)
\label{e:non_local_H}
\end{equation}
with the position dependent, non-local Hamiltonian matrix elements,
\begin{equation}
H_{nm}(\vr,\vr') = \sum\limits_j \sum\limits_{\vk\vk'}\sum\limits_{\vG\vG'} 
\tilde{u}_{n,\vG+\vG_1}^* \tilde{H}_{jm,\vG-\vG'}(\vk) \tilde{u}_{j\vG'} e^{\im(\vk_1\cdot\vr-\vk'\cdot\vr')}
\end{equation} 
Here again we can use the known transformation matrix elements $Q_{n\vG}=\tilde{u}_{n\vG}$.

\subsection{Envelope functions}
With Eq.~(\ref{e:non_local_V}) and Eq.~(\ref{e:non_local_H}) we obtain a unique expansion in terms of $u_n(\vr)$. The envelope functions are then given by a set of non-local envelope functions,
\begin{equation}
  - \frac{\hbar^2}{2m_0} \vnabla^2 f_n(\vr)
  - \frac{\im\hbar}{m_0} \sum\limits_m \vp_{nm} \cdot \vnabla f_m(\vr) 
  + \sum\limits_m \int \rd^3r'\left[ H_{nm}(\vr,\vr') + V_{nm}(\vr,\vr') \right] f_m(\vr')
  = E\, f_n(\vr).
  \label{non_local_envelope_function_equation}
\end{equation}
These equations are much harder to solve due to the non-local interactions and the extra coupling due to the external potential. However, using the eigendecomposition detailed in this paper, the needed elements are all known and the equations are at least solvable in principle. \\[1em]

\section*{References}
\bibliographystyle{elsarticle-num}
\bibliography{main}

\end{document}